\documentclass[%
 aip,
 amsmath,amssymb,
reprint,%
]{revtex4-1}

\usepackage{graphicx}
\usepackage{dcolumn}
\usepackage{bm}

\usepackage[utf8]{inputenc}
\usepackage[T1]{fontenc}
\usepackage{mathptmx}
\usepackage{etoolbox}
\usepackage{comment}
\usepackage{makecell}
\usepackage{float}
\usepackage{bbm}
\usepackage{caption}

\makeatletter
\def\@email#1#2{%
 \endgroup
 \patchcmd{\titleblock@produce}
  {\frontmatter@RRAPformat}
  {\frontmatter@RRAPformat{\produce@RRAP{*#1\href{mailto:#2}{#2}}}\frontmatter@RRAPformat}
  {}{}
}%
\makeatother
\begin{document}

\preprint{AIP/123-QED}

\title[Rb$^+$ and Cs$^+$ SIM]{Rubidium and cesium ion-induced electron and ion signals for scanning ion microscopy applications}
\author{Y. Li}
\author{S.Xu}
\affiliation{%
Department of Applied Physics, Eindhoven University of Technology (TU/e)\\P.O. Box 513, 5600 MB Eindhoven, the Netherlands\\}%
\author{T. H. Loeber}
\affiliation{Nano Structuring Center, Rheinland-Pfälzische Technische Universität Kaiserslautern-Landau (RPTU), Kaiserslautern, Germany}
\author{E. J. D. Vredenbregt\textsuperscript{*}}
 \email{e.j.d.vredenbregt@tue.nl}
\affiliation{%
Department of Applied Physics, Eindhoven University of Technology (TU/e)\\P.O. Box 513, 5600 MB Eindhoven, the Netherlands\\}%

\date{\today}

\begin{abstract}
Scanning ion microscopy applications of novel focused ion beam (FIB) systems based on ultracold rubidium (Rb) and cesium (Cs) atoms were investigated via ion-induced electron and ion yields. Results measured on the Rb$^+$ and Cs$^+$ FIB systems were compared with results from commercially available gallium (Ga$^+$) systems to verify the merits of applying Rb$^+$ and Cs$^+$ for imaging. The comparison shows that Rb$^+$ and Cs$^+$ have higher secondary electron (SE) yields on a variety of pure element targets than Ga$^+$, which implies a higher signal-to-noise ratio can be achieved for the same dose in SE imaging using Rb$^+$/Cs$^+$ than Ga$^+$. In addition, analysis of the ion-induced ion signals reveals that secondary ions dominate Cs$^+$ induced ion signals while the Rb$^+$/Ga$^+$ induced signals contain more backscattered ions.    
\end{abstract}

\maketitle

\section{\label{sec:introduction}Introduction}

Scanning electron microscopy (SEM) and scanning ion microscopy (SIM) have long been used for micro/nano-scale imaging. SIM can offer some advantages over SEM, such as higher surface sensitivity, stronger crystal orientation-based contrast, and decreased susceptibility to sample charging \cite{Ishitani1997SIMvsSEM}. SIM imaging often provides opposite material contrast from SEM, which helps users to view regions of interest that appear dark in SEM imaging \cite{Ohya2007SEM-SIM}. Due to these properties, SIM has applications for materials science such as crystallography mapping \cite{Giannuzzi2005FIBintro}. On the other hand, SIM is intrinsically a destructive technique because it removes material during imaging. In addition to these imaging features, SIM is valuable because it provides real-time monitoring for endpoint detection, which is crucial for semiconductor production and sample preparation \cite{Giannuzzi2005FIBintro}.

Focused ion beam (FIB) microscopes, such as gallium (Ga$^+$) FIB \cite{orloff2003HighResFIB} and helium (He$^+$) FIB\cite{ward2006HeFIB}, have imaging capabilities for SIM. The Ga$^+$ ion is a natural candidate for SIM since most FIB systems use Ga$^+$. However, Ga$^+$ FIBs have drawbacks, including lower brightness at low beam energies and high sample contamination \cite{xiao2013Gaimplant}. These drawbacks lead to inferior spatial resolution and greater beam-induced damages for Ga$^+$ SIM compared to He$^+$ SIM. Conversely, He$^+$ FIB systems typically can deliver high-brightness, low-current beams \cite{orloff2003HighResFIB}. Despite the SIM capabilities, the low current limits He$^+$ FIB in other applications such as sample preparation.  

As alternatives, FIB systems based on laser-cooled lithium (Li) \cite{Twedt2014LiFIB}, rubidium (Rb) \cite{tenhaaf2014ultracoldRBFIB}$^,$ \cite{tenHaaf2018Rbspread}, and cesium (Cs) \cite{Steele2017CsFIB} atoms were developed to deliver a high-brightness beam even at lower beam energies comparable with the Ga$^+$ FIBs. The Li$^+$ FIB has already demonstrated low-energy SIM capabilities \cite{Twedt2014LiFIB} that can significantly reduce sample damage \cite{Brongersma2007lowenergy}$^,$ \cite{Grehl2011LEIS}. Likewise, Rb$^+$ \cite{xu2023Rbdamage} and Cs$^+$ \cite{drezner2016energetic} irradiation on Si has respectively led to lower sample contamination compared to Ga$^+$. For SIM using the Cs$^+$ FIB, Loeber \cite{CsFIBApp} has shown advantages over standard Ga$^+$ FIBs, such as higher resolution, larger depth of focus, and greater material contrast. 

This work aims to study the SIM capability of novel ultracold FIB systems by analyzing the imaging signal yields of element standards under Rb$^+$ or Cs$^+$ irradiation. Firstly, the signals for FIB imaging are introduced in Sec.\ref{sec:imaging-sig}. The FIB imaging mechanism and the data acquisition procedure are then given in Sec.\ref{sec:imaging-mech}. This is followed by Rb$^+$ and Cs$^+$ induced particle yields measured on samples with a large range of atomic numbers ($Z_2$). These measured yields are compared with results from standard Ga$^+$ FIB to demonstrate properties of Rb$^+$/Cs$^+$ SIM, including higher secondary electron (SE) yields and potentially larger material contrast than Ga$^+$ SIM. In addition, secondary ions (SIs) and backscattered ions (BSIs) simulation data are provided to help understand the measured yields.  

\section{\label{sec:imaging-sig}Signals for FIB imaging}

Ion beam irradiation on the sample induces various species of particles, including SEs, SIs, and sputtered neutral atoms. The irradiated sample can also scatter the incident ions elastically to generate BSIs. SIM images are typically formed via the collection of secondary electrons or ions. Only the SE originating within tens of nanometers from the surface can escape the sample to reach the imaging detector. For the image-forming ions, this distance reduces to a few nanometers \cite{Giannuzzi2005FIBintro}. These shallow escape depths guarantee the surface sensitivity of SIM. When ions are collected for SIM imaging, an increased material contrast compared to SEM can be achieved. This contrast results from more significant variations among signal yields from different target materials \cite{Ishitani2002SIMcontrast}. A study of the imaging-forming particle properties, such as signal yields, can help to establish a better understanding of SIM.

During ion bombardment of a target, elastic interactions lead to momentum transfer from the incident ions to the substrate atoms within a collision cascade region. When a surface atom receives more kinetic energy from a collision than the surface binding energy, it can be ejected and become a sputtered particle. Positively or negatively charged SIs are formed when such atoms are ionized during sputtering \cite{Giannuzzi2005FIBintro}. The incident ion can be implanted within the substrate when all its kinetic energy is lost or backscattered to form BSI due to elastic collisions. The SIs and the BSIs differ in their energies and chemical species. The energy of most SIs is less than 50 eV while the energy of the BSI is in the range of the primary ions. The ion species of SIs depends on the target material, while the BSIs always have the same ion species as the primary ions. SI and BSI can be collected for SIM and are usually not differentiated during imaging.

The SEs result from inelastic scattering when electrons from the substrate are ejected by either the incident primary particles (SE\textsubscript{1}) or the exiting secondary or backscattered particles (SE\textsubscript{2}) \cite{Twedt2014LiFIB}. These SEs typically have energies less than 50 eV. The mechanism of ion-induced electron emission is similar to what has been studied for electron-induced electron emission in SEM \cite{Goldstein2018SEM}, except for a higher $\mathrm{SE}_{1}/\mathrm{SE}_{2}$ ratio \cite{Ishitani1997SIMvsSEM}. This increased $\mathrm{SE}_{1}/\mathrm{SE}_{2}$ ratio means the signals for SIM are more localized to the beam-sample interaction area. SE yields are 10-1000$\times$ greater than those for SI or BSI from a typical ion-sample interaction, so SEs are the primary imaging signal for SIM \cite{orloff2003HighResFIB}.   

\section{\label{sec:imaging-mech}Experimental procedure}

\subsection{\label{sec:imaging-modes}FIB imaging modes}

SIM experiments were performed on the in-house developed ultracold Rb$^+$ FIB at TU/e and the commercially available Cs$^+$ FIB by zeroK NanoTech Corporation \cite{zeroKFIB} at RPTU. Detailed descriptions of the Rb$^+$ and Cs$^+$ FIB systems can be found in the works by ten Haaf \cite{tenhaaf2017ultracoldRBFIB} and Knuffman et al \cite{Knuffman2013CeFIB}. For comparison, parallel  Ga$^+$ SIM experiments were conducted on two nearly identical Thermo Fisher Scientific NanoLab\textsuperscript{TM} DualBeam systems \cite{DualBeam} equipped with Ga$^+$ FIB. Both the Rb$^+$ and Cs$^+$ FIB systems have a continuous dynode electron multiplier (CDEM) detector for imaging signal collection. A metal grid collector is installed on the entrance of the CDEM detector facing the ion-sample interaction region. A grid bias is applied to the collector to attract either the positive or the negative ion-induced particles. The Ga$^+$ FIB systems use an Everhart-Thornley detector (ETD) for SE signals and an ion conversion and electron (ICE) detector for SI signals. Despite the difference in detectors, the Ga$^+$ systems deploy a similar bias scheme to attract particles. Inside the CDEM detector is a glass funnel with a wire-spoke structure commonly referred to as the front-end. The biased front-end structure accelerates the attracted particles into the glass funnel. The coating material inside the funnel amplifies the collected signals as the particles travel through the detector. The FIB control software uses the resulting amplified signals to form images. 

Imaging signals are collected as the ion beam scans over commercially available energy dispersive x-ray spectroscopy (EDS) standards made by Ted Pella, Inc. \cite{EDSstandards} under SE or SI mode. The detector is biased positively in the SE mode to attract negative particles (mainly SEs) and negatively in the SI mode to attract positive particles. System default detector bias settings were used during imaging. The detector gain was kept constant for an ion species with a given beam energy. It is worth pointing out that in the SI mode both the positive SIs and BSIs are collected by the CDEM or ICE detector and contribute to forming the images. 

\begin{figure}[h]
    \centering
    \includegraphics[width=0.9\linewidth]{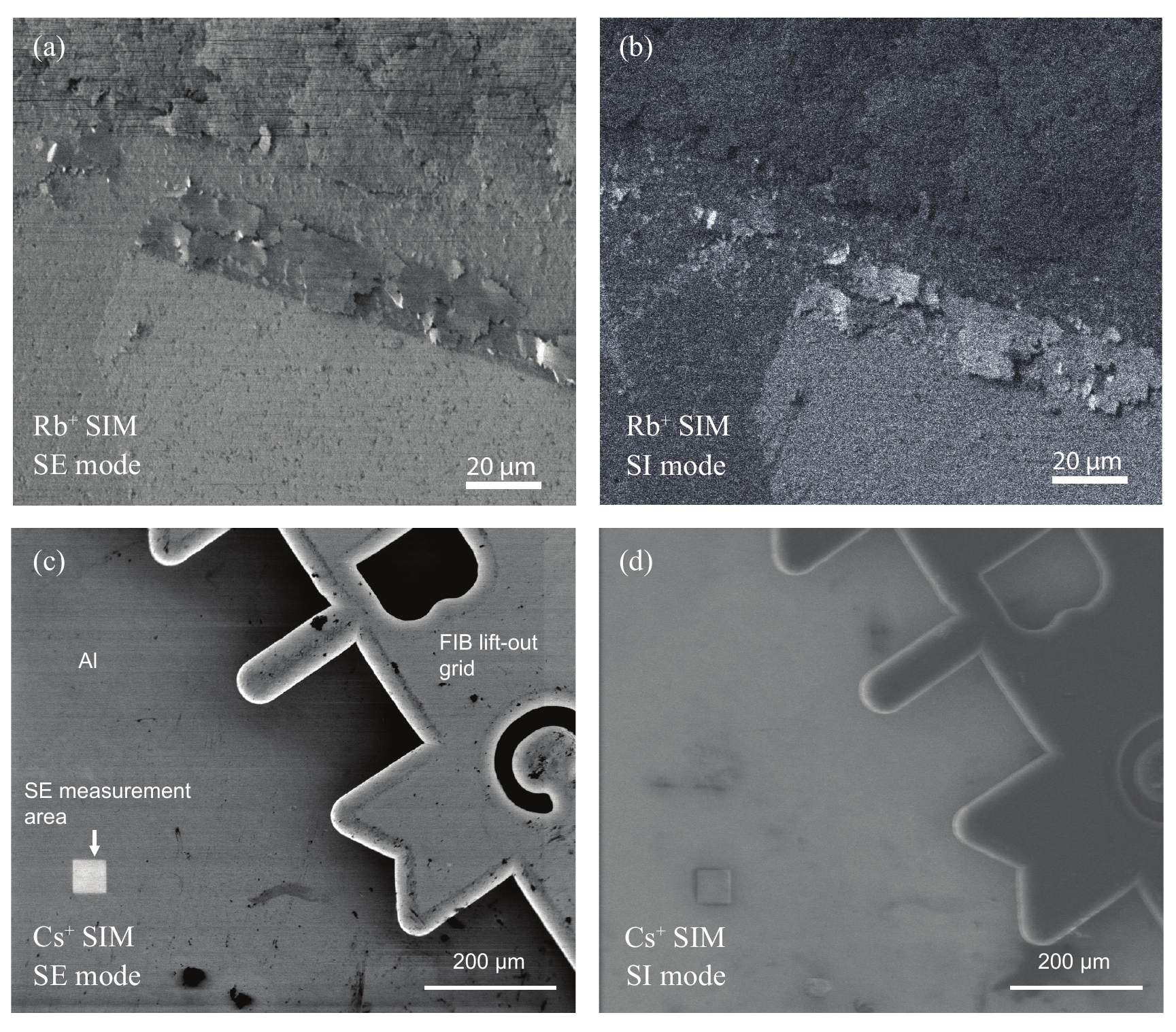}
    \caption{SIM images: Au sample imaged by 8.5 keV 9.8 pA Rb$^+$ in (a) SE mode and (b) SI mode. Al standard, partially covered by a FIB lift-out grid, imaged by 8.0 keV 50 pA Cs$^+$ in (c) SE mode and (d) SI mode. An arrow points to the milled area for SE yield measurement in (c). The slight shift between the SE and SI mode images resulted from the ion beam path shift caused by detector grid biases.}
    \label{fig:SEvsSI}
\end{figure}

Figure \ref{fig:SEvsSI} contains Rb$^+$ and Cs$^+$ SIM images collected in SE and SI mode at the same spot on the samples respectively. Fig.\ref{fig:SEvsSI}(a) and \ref{fig:SEvsSI}(b) show a gold (Au) sample, and Fig.\ref{fig:SEvsSI}(c) and \ref{fig:SEvsSI}(d) an aluminum (Al) standard partially covered by a FIB lift-out grid for transmission electron microscopy (TEM) samples. The difference between SE and SI mode imaging is most apparent when comparing Fig.\ref{fig:SEvsSI}(c) and \ref{fig:SEvsSI}(d). The Al standard within the etched letters "B" and "C" in the FIB lift-out grid appears much darker in the SE image than in the SI image, indicating that SI-mode imaging is less sensitive to shading caused by large surface topographic changes. On the contrary, the Al standard and the FIB lift-out grid have similar gray-scale levels in the SE image while appearing different in the SI image. Such change in gray-scale levels corresponds to a different mechanism of material contrast between SE and SI mode imaging.    

\subsection{\label{sec:SEY measurement}Secondary electron yield measurement}

The absolute SE yields were measured using a setup as described by Chen et al \cite{pchen2009SEinFIBID}. This setup consists of a picoampere current meter connected to the sample stage to record the stage current $I_{\mathrm{stage}}$ as the primary ion current $I_{\mathrm{beam}}$ scans over the sample. Ion beam currents ranged between 10\textendash20 pA for measurements presented in this work. $I_{\mathrm{beam}}$ was first determined using a Faraday cup in the sample stage holder, and then blanked and deflected into another Faraday cup in the FIB column. Current measurements with the two Faraday cups agreed with each other, confirming that all the blanked current can reach the sample during scanning. The blanking mechanism was used to measure all the $I_{\mathrm{beam}}$ data. Since the difference between $I_{\mathrm{stage}}$ and $I_{\mathrm{beam}}$ during ion irradiation is dominated by SEs leaving the sample \cite{orloff2003HighResFIB}, it makes a good quantification of the SE current $I_{\mathrm{SE}}$. Subsequently, the SE yield $\gamma$ can be defined as
\begin{equation}\label{SEY}
    \gamma = \frac{I_{\mathrm{SE}}}{I_{\mathrm{beam}}} = \frac{I_{\mathrm{stage}}-I_{\mathrm{beam}}}{I_{\mathrm{beam}}}.
\end{equation}
During a SE yield measurement, the ion beam scanned continuously over a selected area on the sample until a steady $I_{\mathrm{stage}}$ was reached for data recording. This steady $I_{\mathrm{stage}}$ was used to calculate the corresponding SE yield. A scanned region for SE measurement can be seen in Fig.\ref{fig:SEvsSI}(c). In this SE mode image, the Al surface scanned by Cs$^+$ appears brighter than the area outside of the scanning region.   

\subsection{\label{sec:SIY measurement}SI mode imaging signal measurement}

Signals for SI mode SIM are a combination of positively charged SIs and BSIs. The challenge for quantifying SI mode signals is due to a much smaller SI current ($I_{\mathrm{SI}}$) and BSI current ($I_{\mathrm{BSI}}$) compared to $I_{\mathrm{SE}}$ \cite{orloff2003HighResFIB}. Xu \cite{Xu2023ultracoldRBFIB} conducted a Monte Carlo simulation of ion-solid interactions using the {\small \textsc{SRIM Monte Carlo}} software \cite{biersack1982stopping} to quantify the $I_{\mathrm{BSI}}$/$I_{\mathrm{beam}}$ ratios for metallic substrates irradiated by Rb$^+$. For 8.5 keV Rb$^+$, the maximum $I_{\mathrm{BSI}}$/$I_{\mathrm{beam}}$ ratio is 0.15 for Au target. This $I_{\mathrm{BSI}}$/$I_{\mathrm{beam}}$ ratio significantly decreases for lighter targets (e.g. 0.002 for Cu). A similar trend is seen in Ga$^+$ simulation with the $I_{\mathrm{BSI}}$/$I_{\mathrm{beam}}$ ratio at 0.22 for Au and zero for Cu, both bombarded by 5 keV Ga$^+$ \cite{Giannuzzi2005FIBintro}. It is worth noticing that SRIM cannot predict the charge state of the sputtered atoms or the backscattered ions. Because of the low $I_{\mathrm{SI}}$ and $I_{\mathrm{BSI}}$, the stage current method for the SE yields cannot be used.    

\begin{figure}[h]
    \centering
    \includegraphics[width=0.9\linewidth]{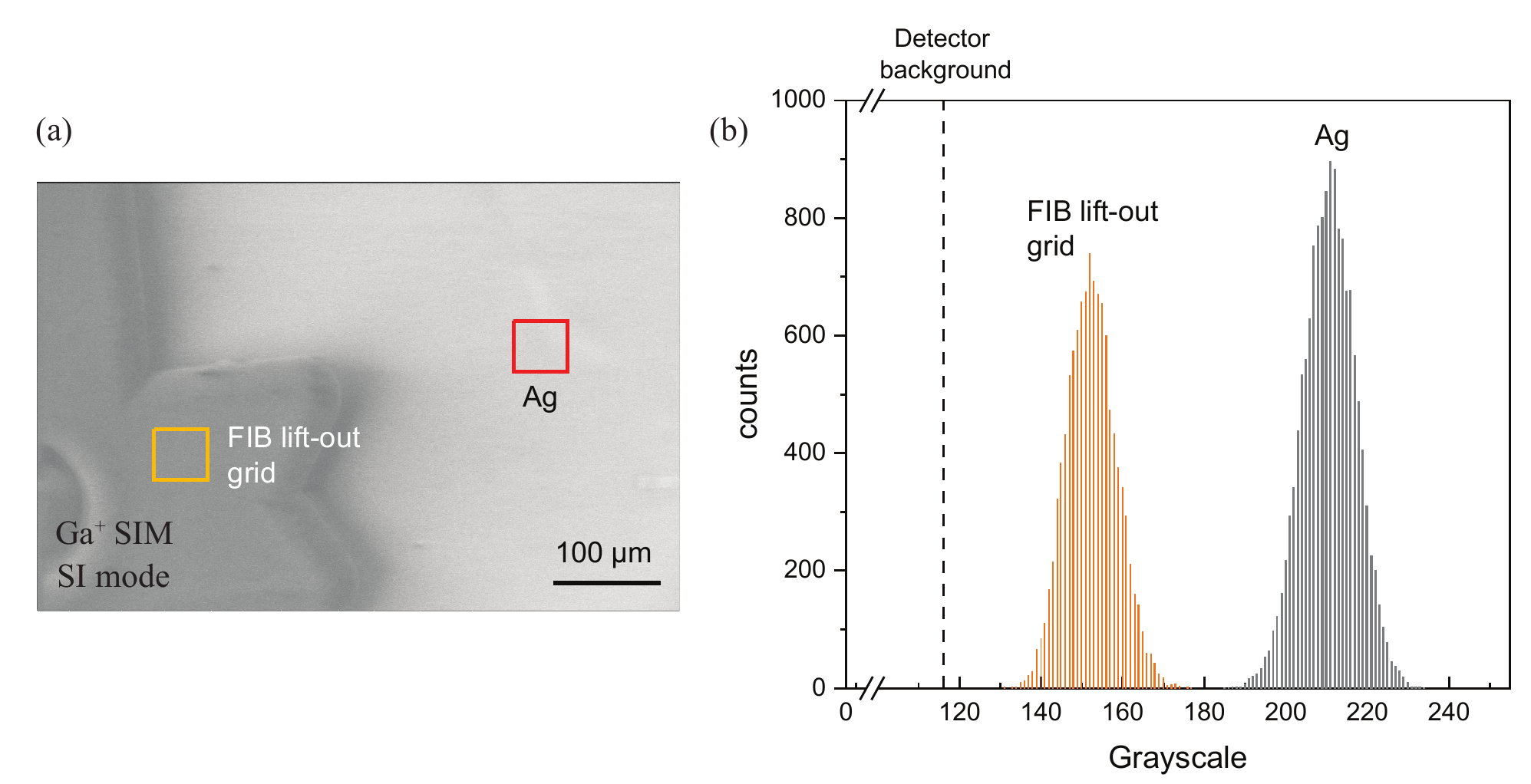}
    \caption{(a) Ga$^+$ SI mode image of a FIB lift-out grid on Ag. The image was made with 8 keV 90 pA Ga$^+$ beam. (b) Histogram of the gray-scale counts of Ag and FIB lift-out grid from the indicated areas in (a).}
    \label{fig:SE measure}
\end{figure}

An indirect way of measuring relative signal yields $Y_{rel}$ of the EDS standards was consequently used by examining their SI mode images. This method is similar to what was proposed by Twedt et al. \cite{Twedt2014LiFIB} for Li$^+$, which compares the gray-scale levels of metal samples to a standard in SI mode images. For each measurement, a low-magnification SI image was acquired of the element to study partially overlaid with a FIB lift-out grid. As an example, Fig.\ref{fig:SE measure}(a) contains a Ga$^+$ SIM SI mode image of a grid on silver (Ag). The mean gray-scale counts in selected areas representing both the element and the FIB lift-out grid were extracted using the image processing software ImageJ \cite{ImageJ}. Figure \ref{fig:SE measure}(b) shows the histogram of the data from Fig.\ref{fig:SE measure}(a). With the mean gray-scale counts of the element target $\overline{GS_{\mathrm{target}}}$ and the FIB lift-out grid $\overline{GS_{\mathrm{Grid}}}$, the relative SI mode imaging signal yield $Y_{rel}$ is defined as       
\begin{equation}\label{Y_rel}
    Y_{rel} = \frac{\overline{GS_{\mathrm{target}}}-\overline{GS_{\mathrm{background}}}}{\overline{GS_{\mathrm{Grid}}}-\overline{GS_{\mathrm{background}}}},
\end{equation}
where $\overline{GS_{\mathrm{background}}}$ is the detector background. This background was acquired from SI mode images when the ion beam was blanked. Measuring $Y_{rel}$ with respect to the FIB lift-out grid circumvents the need to quantify the absolute $I_{\mathrm{BSI}}$ or $I_{\mathrm{SI}}$. Ion scanning changes the $\overline{GS_{\mathrm{target}}}$ to a much lesser degree in SI mode imaging than in SE mode. This effect can be viewed comparing Fig.\ref{fig:SEvsSI}(c) and \ref{fig:SEvsSI}(d) within the SE measurement area. Therefore, unmilled surface areas were used for $Y_{rel}$ calculation.   

\section{\label{sec:results}Results and discussion}

\subsection{\label{sec:SEY}Secondary electron yield}

\begin{figure}[h]
    \centering
    \includegraphics[width=0.95\linewidth]{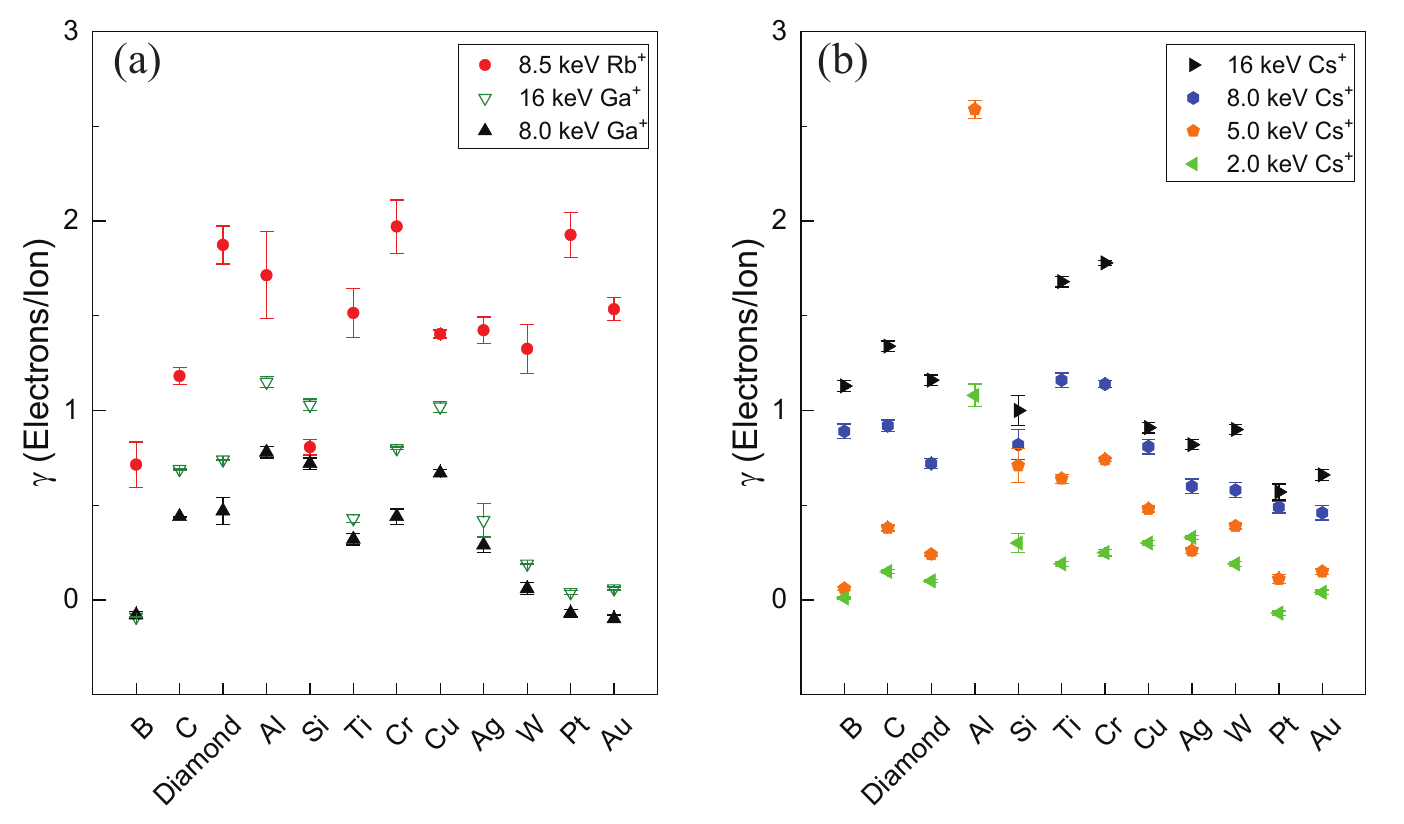}
    \caption{Measured SE yields obtained for element targets irradiated (a) by normal incident 8.0 and 16 keV Ga$^+$ and 8.5 keV Rb$^+$, and (b) by 2.0 - 16 keV Cs$^+$. The SE yield of Al by 8.0 keV and 16 keV Cs$^+$ are $3.24\pm0.08$ and $5.31\pm0.06$, which are not included in to keep the vertical axis ranges the same in Fig.\ref{fig:SEY}(a) and \ref{fig:SEY}(b).}
    \label{fig:SEY}
\end{figure}

Figure \ref{fig:SEY}(a) displays SE yields of a group of EDS standards under 8.5 keV Rb$^+$, 8.0 keV Cs$^+$, compared with 8 keV and 16 keV Ga$^+$ irradiation. Figure \ref{fig:SEY}(b) contains SE yields of the same targets by 2-16 keV Cs$^+$ irradiation. All measurements were made at normal ion incidence with sample chamber pressure in the low $10^{-6}$ mbar. The error bars represent the standard deviation of the yields and are dominated by the fluctuation in the stage current. To investigate the effects of different FIB systems on the measurements, 8 keV Ga$^+$ SE yields were measured on the same EDS standards in the two DualBeam systems mentioned in Sec. \ref{sec:imaging-modes}. The outcome shows less than 10\% differences between the two data sets. The small differences indicate that the SE yields are more affected by primary ion species than by the individual FIB systems. 

One trend in the data is that the SE yield increases with incident ion beam energy, as can be seen for Ga$^+$ in Fig.\ref{fig:SEY}(a) and Cs$^+$ in \ref{fig:SEY}(b). This trend agrees with previous findings for Ga$^+$ on clean metal targets \cite{Ohya2007SEM-SIM} $^,$ \cite{Lakits1990SEmetal}. A higher incident beam energy carries more kinetic energy that can be transferred to the target electrons for SE emission. Ishitani et al. have also found that for 30 keV Ga$^+$ $\gamma$ decreases with target atomic number ($Z_2$) beyond copper (Cu) \cite{Ishitani2002SIMcontrast}. After colliding with ions with the same kinetic energy, high-$Z_2$ target atoms reach a slower velocity. Therefore, these atoms transfer less energy to the excited electrons. The reduced transferable energy for excitation also decreases the multiplication of other excited electrons within the collision cascade. This $\gamma$-$Z_2$ correlation is less obvious in the Rb$^+$ or the Cs$^+$ data. Fine structures exist within the $\gamma$ data, such as the local maxima near Al and Cu among the Ga$^+$ data. Such local maxima are likely due to variation in the target metal work function \cite{Suzuki2008Materialcontrast}. However, the exceptionally high $\gamma$ for Al by Cs$^+$ is likely due to additional mechanisms that are yet to be investigated. 

The higher atomic mass of the incident ion ($M_1$) can partially explain why the SE yields of Rb$^+$ and Cs$^+$ are consistently higher than for Ga$^+$. As Ohya and Ishitani state, ion-induced SEs can be produced by three mechanisms: excitation by primary ions, by recoiled material atoms, and by electron cascade \cite{Ohya2007SEM-SIM}. Ishitani et al. \cite{Ishitani2002SIMcontrast} postulate that the surface barrier for excited electrons sees a larger reduction when heavier ions bombard target surfaces. This reduction in the surface barrier makes it easier for the excited electrons to escape from the target. Another possible explanation is the higher amount of SE emission from recoiling target atoms caused by Rb$^+$ or Cs$^+$ than Ga$^+$ \cite{Alonso1980SEYZ1}. Alonso et al. \cite{Alonso1980SEYZ1} also partially attribute the increase in $\gamma$ of heavier primary ions to the decrease in the mean free path. Heavier primary ions, with a small mean free path, spend more energy within the electron escape depth. This also makes excited electrons more likely to leave the target. 

Nonetheless, the explanation given above does not account for the SE yields of Cs$^+$ being lower than those of Rb$^+$ despite having a higher $M_1$. To explain the change in SE yields with respect to the incident ion species ($Z_1$) requires a deeper look into the ion-induced SE emission mechanism. Ion-induced SEs are created in two processes: potential and kinetic emission \cite{Baragiola1993ISEmechanisms}. In potential emission, SEs are excited by the neutralization of the incident ions, while in kinetic emission, SEs are generated by kinetic energy transfer from the incident ions to the substrate. An empirical model for potential emission yield is 
\begin{equation}\label{PSEY}
    \gamma_P = 0.032(0.78E_i-2\Phi),
\end{equation}
where $E_i$ is the ionization potential of the primary ion and $\Phi$ the work function of the substrate surface in electronvolts \cite{Baragiola1979ISEfromMetals}. Research on the $Z_1$ dependence of kinetic SE emission has found that Cs$^+$ can induce more SEs on Au, Ag, and Pd targets and fewer SEs on Cu target than Ga$^+$ or Rb$^+$ \cite{Ferguson1989Z1ISE}. However, this measurement was conducted for thin targets (thickness $<$ a few nm) with low primary ion currents. Such an experimental setup provided a different environment for ion accumulation in the sample from the FIB-SIM study presented in this work. Instead, potential emission can perhaps explain our observation. Since the $\gamma$ data were collected when the stage current stabilized, the accumulated ions could lower the surface work function $\Phi$, which can increase the potential emission yield as predicted by Eq. \ref{PSEY}. One example of the accumulated ions changing the SE emission could be viewed in Fig.\ref{fig:SEvsSI}(c). The indicated SE measurement area, having been milled by Cs$^+$, appears brighter than the unmilled Al surface. This change in the gray-scale level means a higher SE yield within the milled Al post Cs$^+$ accumulation.   

Although only speculations about Rb$^+$ and Cs$^+$ SE yield behavior are given, the data suggest a potential advantage of using Rb$^+$/Cs$^+$ over Ga$^+$ for SIM. Higher SE yields can help to obtain a higher signal/noise ratio for imaging. This is especially valuable for accurate real-time monitoring or endpoint detection during sample preparation. 

\subsection{\label{sec:imaging-SIY}SI mode imaging signal}

\begin{figure}[h]
    \centering
    \includegraphics[width=0.95\linewidth]{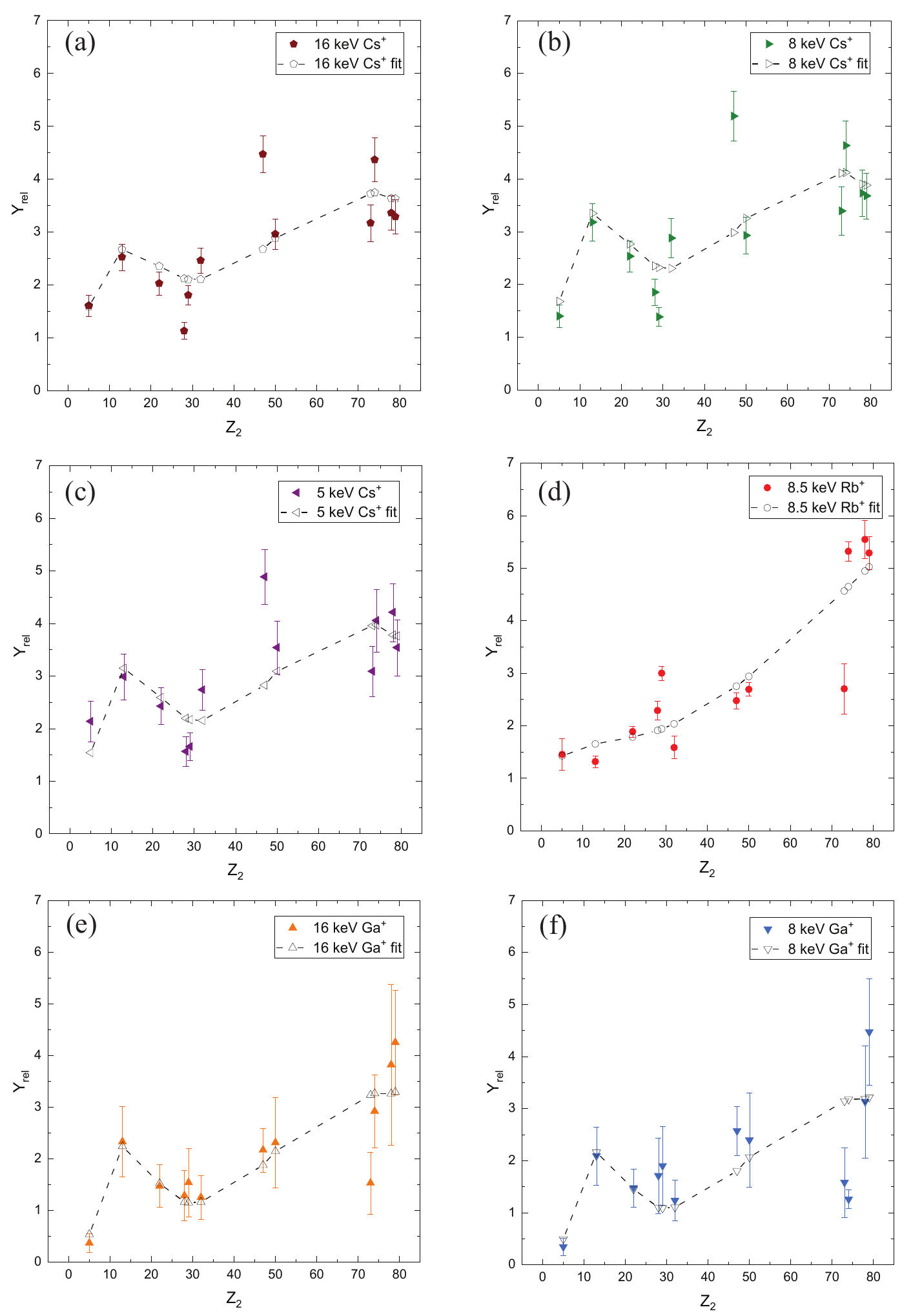}
    \caption{Measured and fitted $Y_{rel}$ for (a) 16 keV, (b) 8 keV, (c) 5 keV Cs$^+$, (d) 8.5 keV Rb$^+$, and (e) 16 keV, (f) 8 keV Ga$^+$. The measured $Y_{rel}$'s are indicated by filled markers and the fitted $Y_{rel}$ by unfilled markers joined by a dashed line as a guide to the eye. }
    \label{fig:Ga Rb fit}
\end{figure}

Figure \ref{fig:Ga Rb fit} exhibits the measured $Y_{rel}$ for twelve EDS standard targets irradiated by Cs$^+$ (Fig.\ref{fig:Ga Rb fit}(a)-(c)), Rb$^+$ (Fig.\ref{fig:Ga Rb fit}(d)), and Ga$^+$ (Fig.\ref{fig:Ga Rb fit}(e)-(f)). The error bars on the measured data were calculated by propagating the standard deviations of the gray-scale counts. In general, $Y_{rel}$ rises with increasing $Z_2$ for each primary ion species for beam energies. Local maxima in $Y_{rel}$ values appear near $Z_2 = 29$, $Z_2 = 47$, and $Z_2 = 79$, which agrees with the measured relative BSI yields reported for He$^+$ \cite{Sijbrandij2008HeBSI} and Li$^+$ \cite{Twedt2014LiFIB}. Local $Y_{rel}$ maxima also appear near Al ($Z_2 = 13$) in the Cs$^+$ and Ga$^+$ data. Twedt et al. \cite{Twedt2014LiFIB} speculate that these $Y_{rel}$ peaks correlate to the changes in the electronic stopping powers of the targets. 

A simple fitting model was constructed to understand the relative contribution of positive SI and BSI to $Y_{rel}$. The positive ion yield $n^+$ of a sputtered element can be described by  
\begin{equation}\label{SI model}
    n^+ \propto \exp{(-\beta I)}, 
\end{equation}
where $I$ is the free space ionization potential of the sputtered atom and $\beta$ is a constant \cite{Anderson1972SIY}. The BSI yield $\sigma$ is proportional to the standard Rutherford scattering cross section   
\begin{equation}\label{BI model}
    \sigma \propto (Z_1 Z_2/E)^2,
\end{equation}
where $E$ is the incident beam energy \cite{Twedt2014LiFIB}. Then $Y_{rel}$ can be divided into the SI and the BSI contribution described by 
\begin{equation}\label{fit model}
    Y_{rel} = C_{SI}\exp{(-\beta I)} + C_{BSI}(Z_1 Z_2/E)^2,
\end{equation}
in which $C_{SI}$ and $C_{BSI}$ are two constants. The measured $Y_{rel}$ data were fitted in Microsoft\textsuperscript{\textregistered} Excel\textsuperscript{\textregistered} by this linear combination model using a least-squares algorithm with three fitting coefficients: $\beta$, $C_{SI}$, and $C_{BSI}$. The $Y_{rel}$ from fitting, indicated by unfilled markers joined by a dashed line in Fig.\ref{fig:Ga Rb fit}, are superimposed on the measured $Y_{rel}$'s. The free space ionization potential values for $I$ used for fitting were acquired from the National Institute of Standards and Technology database \cite{NISTdata}.

\begin{figure}[h]
    \centering
    \includegraphics[width=0.96\linewidth]{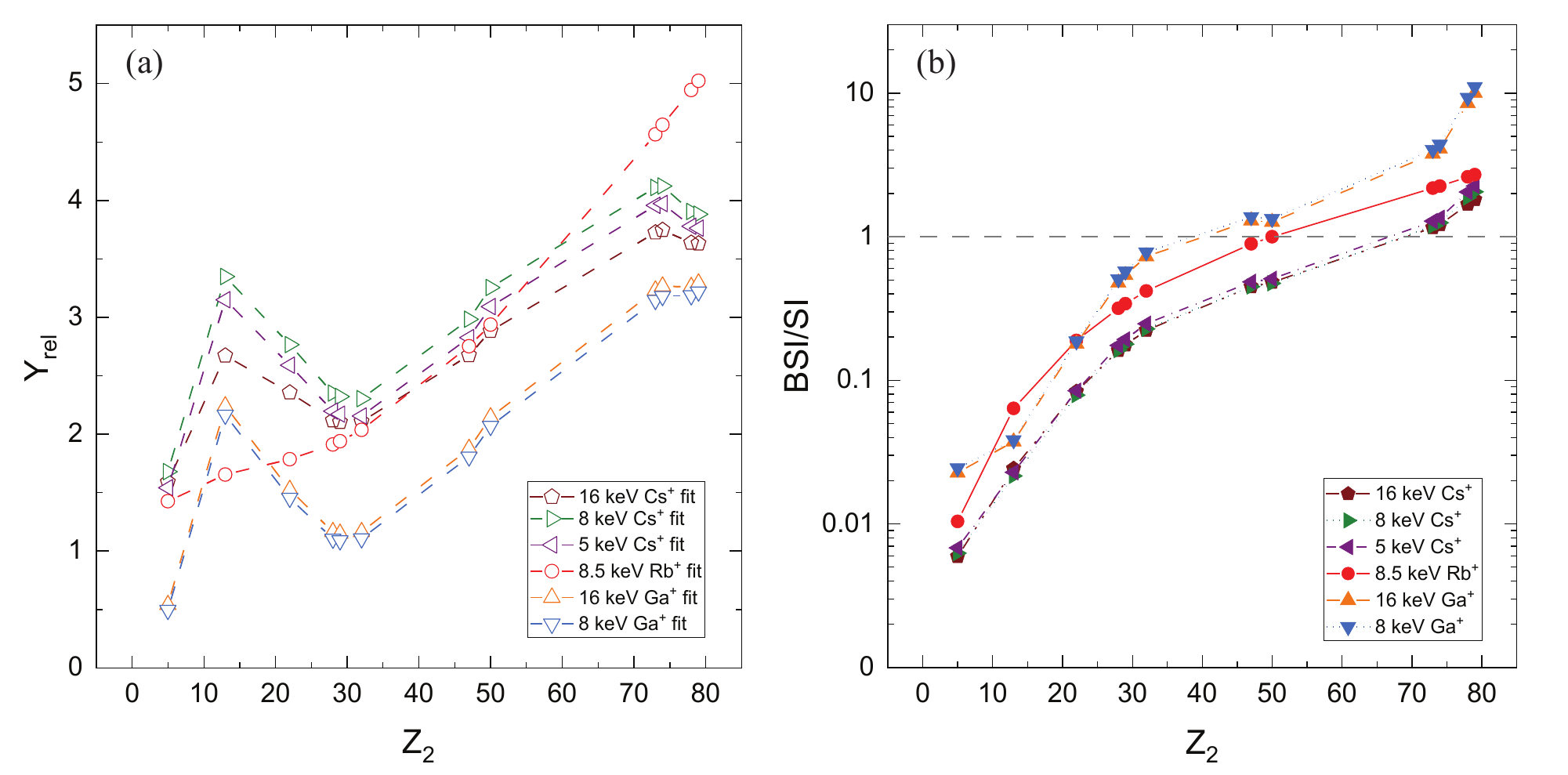}
    \caption{(a) Fitted $Y_{rel}$ values predicted by the linear fitting model for Ga$^+$, Rb$^+$, and Cs$^+$. (b)Ratios of the BSI and SI contribution to the $Y_{rel}$ for Ga$^+$, Rb$^+$, and Cs$^+$ based on the fitted results.}
    \label{fig:BI-SI ratio}
\end{figure}

Based on visual inspection of Fig.\ref{fig:Ga Rb fit}, the predicted $Y_{rel}$ values by fitting lie close to the measured data. For the Cs$^+$ and Ga$^+$ data, the model can describe the local maxima near Al ($Z = 13$) due to the lower ionization potential. The fitting model can also predict the rise of $Y_{rel}$ with $Z_2$. Such a trend can be explained by a growing BSI contribution with increasing $Z_2$. For the Rb$^+$ data, the fitted results mostly follow a parabolic increase with $Z_2$ due to a lack of local maximum at Al. All the fitted results are plotted in Fig.\ref{fig:BI-SI ratio}(a). The predicted $Y_{rel}$ values for Cs$^+$ and Ga$^+$ follow a similar trend with a vertical shift while the $Y_{rel}$ values for Rb$^+$ exhibit a monotonic growth with $Z_2$. These predicted $Y_{rel}$ values imply that SI mode imaging by Cs$^+$ and Ga$^+$, similar to each other, would have different material contrast from imaging by Rb$^+$.

Different ion sources can induce different $\mathrm{BSI}/\mathrm{SI}$ ratios. For example, Li$^+$, lighter than Ga$^+$, is expected to induce predominately BSIs \cite{Twedt2014LiFIB}. Using the model described by Eq. \ref{fit model}, $Y_{rel}$ is divided into a BSI and an SI contribution to calculate the $\mathrm{BSI}/\mathrm{SI}$ ratios caused by different primary ions. Figure \ref{fig:BI-SI ratio}(b) shows the BSI/SI ratios for all the fitted data, which shows that $Y_{rel}$ goes from SI-dominating at lower $Z_2$ to BSI-dominating at higher $Z_2$. The cross-over happens where $\mathrm{BSI}/\mathrm{SI} = 1$. This cross-over shifts to higher $Z_2$ for increasing $Z_1$. The heavier incident ions make SI creation more probable, especially for lower $Z_2$ targets. From Eq. \ref{fit model}, the BSI contribution is predicted to decrease with increasing $E$. In contrast, a higher $E$ leads to more sputtering \cite{Giannuzzi2005FIBintro}, which creates more SIs. The combined effects reduce the BSI/SI ratio for higher $E$. In addition, Cs$^+$ has the lowest BSI/SI values for all $Z_2$, indicating that Cs$^+$-induced ion signals are mainly SI-dominated. This agrees with the high SI yields of Cs$^+$ from secondary-ion mass spectrometry (SIMS) studies \cite{storms1977CsSIM}. 

This fit model using the linear combination of SI and BSI contribution still has some limitations. According to Eq. \ref{fit model}, the BSI contribution grows with $Z_2^2$ monotonously. Deviations from this behavior can only be generated by the SI contribution via variation in ionization potentials. Although the model can well fit the peaks at Al for Ga$^+$ $Y_{rel}$, it does not fit some of the other peaks such as the ones near $Z_2 = 47$. Besides ionization potentials, these peaks might be due to sample surface conditions, influence of free atoms \cite{Williams1980SImodel}, or high sputter yields. A possible improvement for the linear combination model is to add a material dependency of $C_{SI}$. 

\section{\label{sec:conclusion}Conclusion}

Scanning ion microscopy was performed for ultracold Rb$^+$ and Cs$^+$ FIB systems and benchmarked against data from comparison experiments conducted on Ga$^+$ FIBs for both SE and SI mode imaging. The SE yields for pure elements reveal that Rb$^+$ and Cs$^+$ ions can induce more SEs compared with Ga$^+$ at similar beam energies. The high SE yields suggest that better imaging quality can be achieved in terms of signal/noise ratio for the same dose. Relative SI imaging signal $Y_{rel}$ results imply similar material contrast for Cs$^+$ and Ga$^+$ and a different one for Rb$^+$ in SI mode imaging. A simple linear model was used to fit the measured $Y_{rel}$ data, which predicts that Cs$^+$ ions can induce higher relative SI contributions than Rb$^+$ and Ga$^+$ under similar ion beam energies. 

\begin{acknowledgments}
This work is part of the project Next-Generation Focused Ion Beam (NWO-TTW16178) of the research program Applied and Engineering Sciences (TTW), which is (partly) financed by the Dutch Research Council (NWO). The authors are (partly) members of the FIT4NANO COST Action CA19140. 
\end{acknowledgments}

\section*{\label{sec:declarations}Author Declarations}
\section*{Conflict of interest}
The authors have no conflicts to disclose.

\section*{Data Availability Statement}
The data that support the findings of this study are available from the corresponding author upon reasonable request.

\nocite{*}

%

\end{document}